\documentclass[letter]{aa}
\usepackage{natbib}
\usepackage{graphicx}
\usepackage[varg]{txfonts}

\begin{document}

\title{On binary driven hypernovae and their nested late X-ray emission}
\titlerunning{On binary driven hypernovae and their nested late X-ray emission}
\authorrunning{Ruffini et al.}
\author{R. Ruffini\inst{1,2,3,4}, M. Muccino\inst{1,2}, C.~L. Bianco\inst{1,2}, M. Enderli\inst{1,3}, L. Izzo\inst{1,2}, M. Kovacevic\inst{1,3},  A.~V. Penacchioni\inst{4},\\ G.~B. Pisani\inst{1,3}, J.~A. Rueda\inst{1,2,4}, Y.~Wang\inst{1,2}}
\institute{Dip. di Fisica and ICRA, Sapienza Universit\`a di Roma, Piazzale Aldo Moro 5, I-00185 Rome, Italy, \and ICRANet, Piazza della Repubblica 10, I-65122 Pescara, Italy, \and Universit\'e de Nice Sophia Antipolis, CEDEX 2, Grand Ch\^{a}teau Parc Valrose, Nice, France, \and ICRANet-Rio, Centro Brasileiro de Pesquisas Fisicas, Rua Dr. Xavier Sigaud 150, Rio de Janeiro, RJ, 22290-180, Brazil.}

\abstract
{
The induced gravitational collapse (IGC) paradigm addresses the very energetic ($10^{52}$--$10^{54}$ erg) long gamma-ray bursts (GRBs) associated to supernovae (SNe).
Unlike the traditional ``collapsar'' model, an evolved FeCO core with a companion neutron star (NS) in a tight binary system is considered as the progenitor.
This special class of sources, here named ``binary driven hypernovae'' (BdHNe), presents a composite sequence composed of four different episodes with precise spectral and luminosity features.
}
{
We first compare and contrast the steep decay, the plateau, and the power-law decay of the X-ray luminosities of three selected BdHNe (GRB 060729, GRB 061121, and GRB 130427A). Second, to explain the different sizes and Lorentz factors of the emitting regions of the four episodes, for definiteness, we use the most complete set of data of GRB 090618. Finally, we show the possible role of r-process, which originates in the binary system of the progenitor.
}
{
We compare and contrast the late X-ray luminosity of the above three BdHNe.
We examine correlations between the time at the starting point of the constant late power-law decay $t^\ast_a$, the average prompt luminosity $\langle L_{iso} \rangle$, and the luminosity at the end of the plateau $L_a$.
We analyze a thermal emission ($\sim0.97$--$0.29$ keV), observed during the X-ray steep decay phase of GRB 090618.
}
{
The late X-ray luminosities of the three BdHNe, in the rest-frame energy band $0.3$--$10$ keV, show a precisely constrained ``nested'' structure.
In a space-time diagram, we illustrate the different sizes and Lorentz factors of the emitting regions of the three episodes.
For GRB 090618, we infer an initial dimension of the thermal emitter of $\sim7\times10^{12}$ cm, expanding at $\Gamma\approx2$.
We find tighter correlations than the Dainotti-Willingale ones.
} 
{
We confirm a constant slope power-law behavior for the late X-ray luminosity in the source rest frame, which may lead to a new distance indicator for BdHNe. These results, as well as the emitter size and Lorentz factor, appear to be inconsistent with the traditional afterglow model based on synchrotron emission from an ultra-relativistic ($\Gamma \sim 10^2$--$10^3$) collimated jet outflow. We argue, instead, for the possible role of r-process, originating in the binary system, to power the mildly relativistic X-ray source.
}


\offprints{\email{marco.muccino@icra.it}}

\date{}

\maketitle

\section{Introduction}

The induced gravitational collapse (IGC) paradigm has been widely illustrated \citep{2006tmgm.meet..369R,Ruffini2007b,2008mgm..conf..368R,IGC,IGC2}. 
It assumes that long, energetic ($10^{52}$--$10^{54}$ erg) gamma-ray bursts (GRBs) associated to supernovae (SNe) originate in a close binary system composed of an evolved massive star (likely a FeCO core) in the latest phases of its thermonuclear evolution and a neutron star (NS) companion.
From an observational point of view, the complete time sequence of the IGC paradigm binary system has been identified in GRB 090618 \citep{Izzo2012}, GRB 101023 \citep{Penacchioni2011}, GRB 110907B \citep{Penacchioni2013}, and GRB 970828 \citep{Rees_nuovo}.
We name these especially energetic systems, here, fulfilling the IGC paradigm, ``binary driven hypernovae'' (BdHNe), to differentiate them from the traditional less energetic hypernovae.

\begin{figure}
\centering
\includegraphics[width=0.7\hsize,clip]{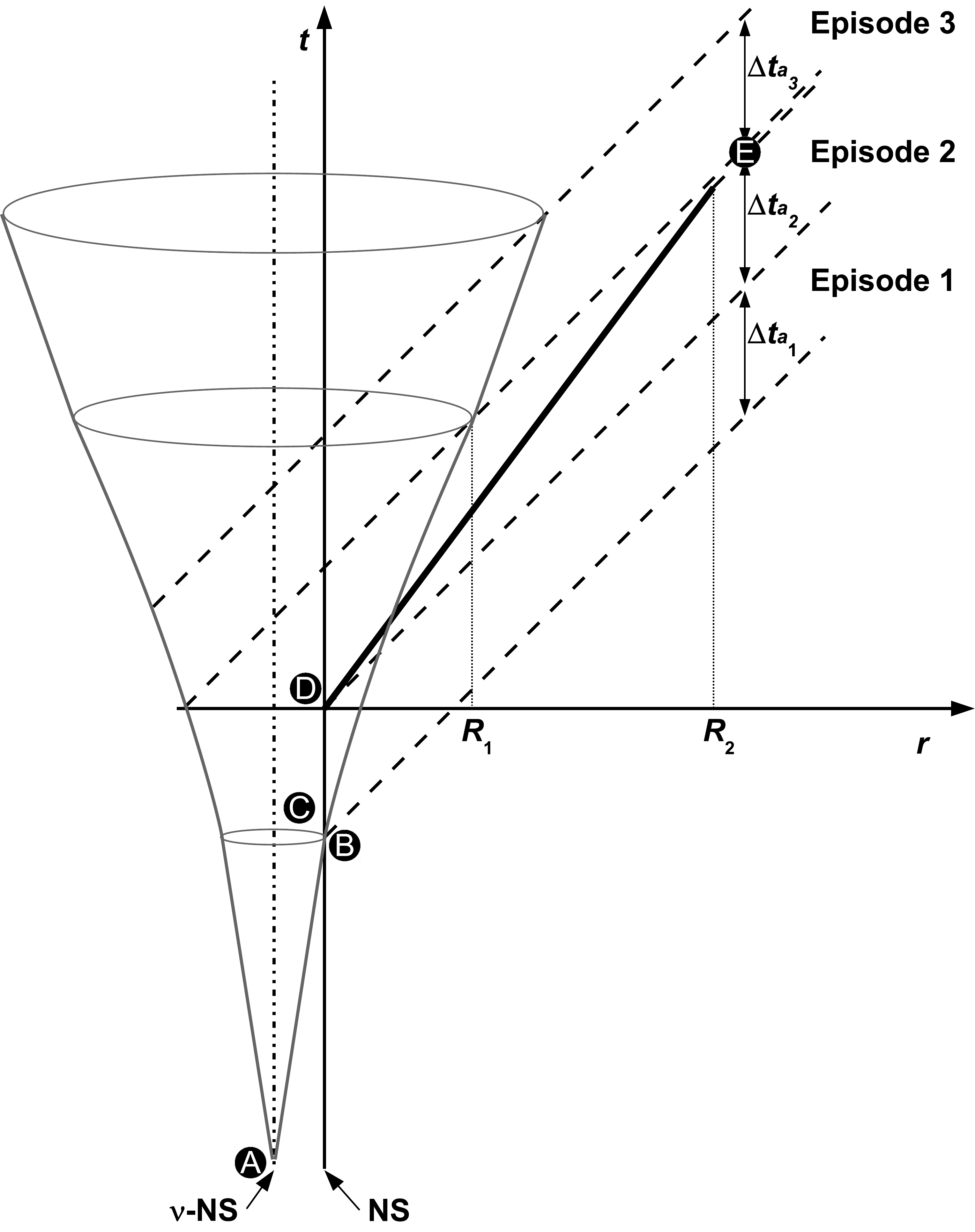}
\caption{IGC space-time diagram (not in scale) illustrates the relativistic motion of Episode 2 ($\Gamma\approx500$, thick line) and the non-relativistic Episode 1 ($\Gamma\approx1$) and Episode 3 ($\Gamma\approx2$). Emissions from different radii, $R_1$ ($\sim10^{13}$ cm) and $R_2$ ($\sim10^{16}$--$10^{17}$ cm), contribute to the transition point (E). Clearly, the X-ray luminosity originates in the SN remnant or in the newly-born BH, but not in the GRB.}
\label{schema}
\end{figure}

In this Letter we introduce the IGC paradigm space-time diagram for the four distinct emission episodes (see Fig. \ref{schema}):\\
\textbf{Episode 1} corresponds to the onset of the FeCO core SN explosion, creating a new-NS ($\nu$-NS, see A).
Part of the SN ejecta triggers an accretion process onto the NS companion \citep[see][and B in Fig. \ref{schema}]{IGC,IGC2}, and a prolonged interaction between the $\nu$-NS and the NS binary companion occurs (C).
This leads to a spectrum with an expanding thermal component plus an extra power law (see Fig. 16 in \citealp{Izzo2012}, and Fig. 4 in \citealp{Rees_nuovo}).\\
\textbf{Episode 2} occurs when the companion NS reaches its critical mass and collapses to a black hole (BH), emitting the GRB (D) with Lorentz factors $\Gamma\approx10^2$--$10^3$ \citep[for details, see e.g.][]{PhysRep,Izzo2012,Rees_nuovo}.\\
\textbf{Episode 3}, observed in the X-rays, shows very precise behavior consisting of a steep decay, starting at the end point of the prompt emission (see E), and then a plateau phase, followed by a late constant power-law decay \citep[see, e.g.,][]{Izzo2012,Penacchioni2011,Rees_nuovo}.\\
\textbf{Episode 4}, not shown in Fig. \ref{schema}, corresponds to the optical SN emission due to the $^{56}$Ni decay \citep[see][]{1996SSRv...78..559A} occurring after $\sim 10$--$15$ days in the cosmological rest frame. In all BdHNe, the SN appears to have the same luminosity as in the case of SN 1998bw \citep{Amati2007}. Although the presence of the SN is implicit in all the sources fulfilling the IGC paradigm, it is only detectable for GRBs at $z\lesssim1$, in view of the limitations of the current optical telescopes.

We are going to see in this Letter that Episodes 1 and 2 can differ greatly in luminosity and timescale from source to source, while we confirm that in Episode 3, the late X-ray luminosities overlap: they follow a common power-law behavior with a constant slope in the source rest frame \citep{Pisani2013}. 
We point out here that the starting point of this power-law component is a function of the GRB isotropic energy $E_{iso}$.

The main goals of this Letter consist in a) comparing and contrasting the steep decay, the plateau, and the power-law decay of the X-ray luminosities as functions of $E_{iso}$ by considering three selected GRBs (060729, 061121, and 130427A); b) pointing out the difference in the size and the Lorentz factors of the emitting regions of Episodes 1, 2, and 3 (for definiteness we use as prototype the source with the most complete dataset, GRB 090618); c) drawing attention to the possible role of the r-process, originating in the binary system of the progenitor, to power the mildly relativistic X-ray emission in the late phases of Episode 3.

\section{The case of GRB 090618}

We illustrate the difference in the emitting region sizes in the three episodes and their corresponding Lorentz factors:\\
\textbf{Episode 1} has a thermal component expanding from $\sim10^9$ cm to $\sim10^{10}$ cm on a rest-frame timescale of $\sim30$ s with an average velocity of $\sim4\times10^8$ cm s$^{-1}$ \citep[see][]{Izzo2012}. 
The total energy is $4.1\times10^{52}$ erg, well above the traditional kinetic energy expected in the early phases of a SN, and it originates in the accretion of the material of the SN ejecta on the companion NS in the binary system \citep{IGC,Rees_nuovo}.\\
\textbf{Episode 2} has been shown to be the ultra-relativistic prompt emission episode (e.g., the actual GRB) stemming from the collapse of the NS to a BH.
Its isotropic energy is $2.5\times10^{53}$ erg.
The characteristic Lorentz factor at the transparency of the fireshell has been found to be $\Gamma=490$ for GRB 090618.
The characteristic spatial extension goes all the way up to  $\sim10^{16}$--$10^{17}$ cm, reached at the end of Episode 2 \citep[see Fig. 10 in][]{Izzo2012}.\\
\textbf{Episode 3} has an isotropic energy of $\approx6\times10^{51}$ erg.
A striking feature occurs during its steep decay phase: in the early observed $150$ s, \citet{Page2011} have found a thermal component with a decreasing temperature from $\sim0.97$ keV to $\sim0.29$ keV \citep[see also][]{Starling2012}.
The surface radius of the emitter can be inferred from the observed temperature $T_o$ and flux $F_{BB}$ of the thermal component. 
We have, in fact \citep{Izzo2012},
\begin{equation}
\label{raggiorel} r\approx\Gamma\, d_l\, (1+z)^{-2} \sqrt{F_{BB}/(\sigma T_o^4)}\ ,
\end{equation}
where $d_l$ is the luminosity distance in the $\Lambda$CDM cosmological model and $\sigma$ the Stefan-Boltzmann constant.
As usual, $\Gamma=1/\sqrt{1-\beta^2}$, where $\beta=v/c$ is the expansion velocity in units of the speed of light $c$.

\begin{figure}
\centering
\includegraphics[width=0.7\hsize,clip]{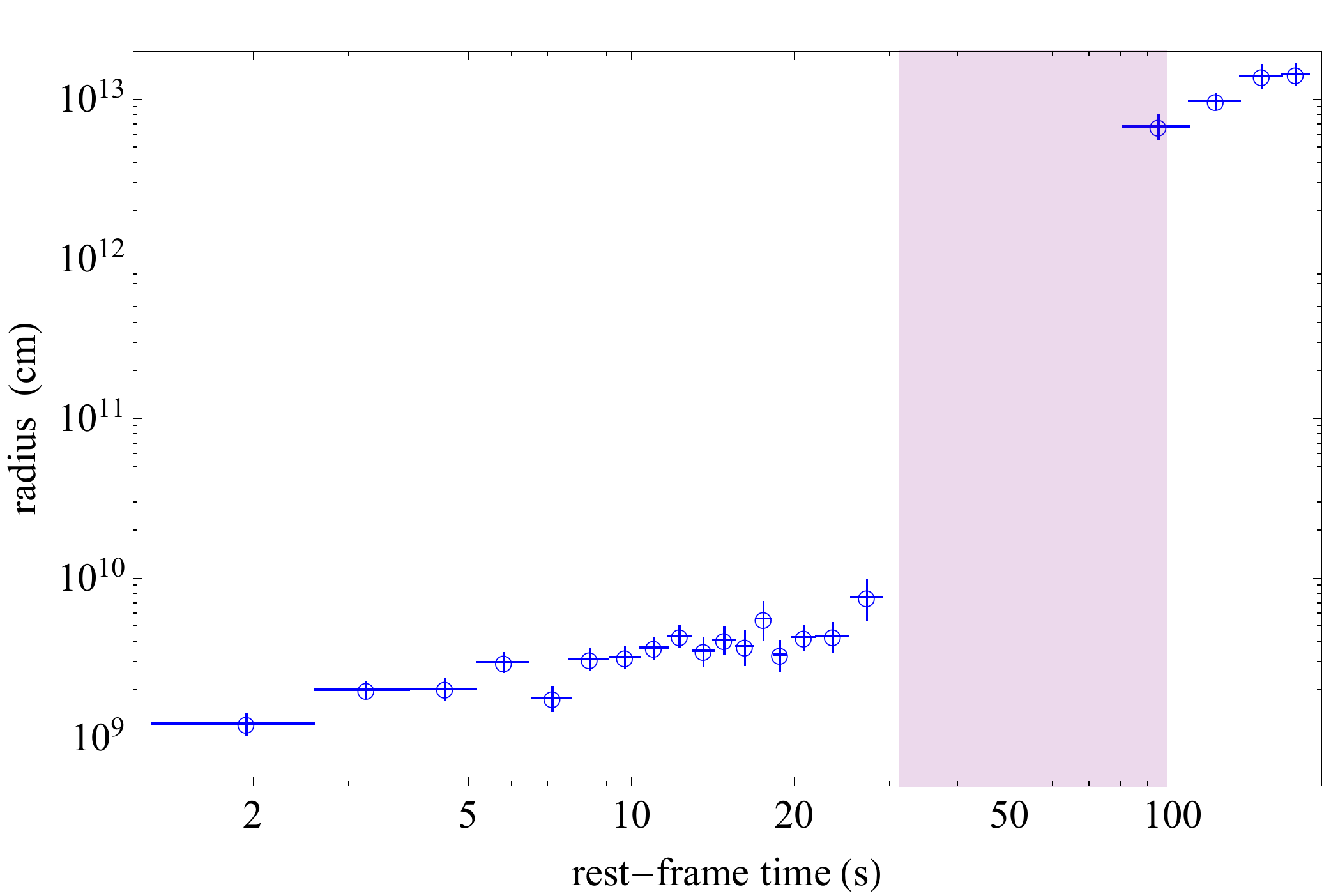}
\caption{Radii (open blue circles) of the emitting regions, measured in the cosmological rest frame. Episode 1 radius ranges from $\sim10^9$ cm to $\sim10^{10}$ and expands at $\Gamma\approx1$ \citep{Izzo2012}. The Episode 3 radius, in the early phases of the steep decay, starts from a value of $\sim7\times10^{12}$ cm and expands at $\Gamma\approx2$. The Episode 2 rest-frame duration is indicated by the shaded purple region. The expansion velocity at late times is expected to approach the asymptotic value of $0.1c$ observed in the optical spectra \citep{DellaValle2011}, in the absence of any further acceleration process.}
\label{rad_ind_tot}
\end{figure}

In parallel, the relation between the detector arrival time $t_a^d$, the cosmological rest-frame arrival time $t_a$ and the laboratory time $t$, is given by $t_a^d\equiv t_a(1+z)=t (1-\beta\cos\theta)(1+z)$, where $\theta$ is the displacement angle of the considered photon emission point from the line of sight \citep[see, e.g.,][]{Bianco2001}.
We can then deduce the expansion velocity $\beta$, assumed to be constant, from the ratio between the variation of the emitter radius $\Delta r$ and the emission duration in laboratory frame $\Delta t$, i.e. $\beta=\Delta r/(c \Delta t)$.
Using the condition $\beta\leq\cos\theta\leq1$ \citep{Bianco2001}, we obtain $0.75\leq\beta\leq0.89$ and, correspondingly, $1.50\leq\Gamma\leq2.19$ and radii $r\sim10^{13}$ cm (see Fig. \ref{rad_ind_tot}).

As is clear from Fig. \ref{schema}, a sharp transition occurs between the end of Episode 2, where the characteristic dimensions reached by the GRB are $\sim10^{16}$--$10^{17}$ cm, and the emission at the beginning of X-ray luminosity, with an initial size of $\sim7\times10^{12}$ cm.
This leads to the conclusion that the X-ray emission of Episode 3 originates in the SN ejecta or in the accretion on the newly born BH and, anyway, not from the GRB.

\section{The ``nested'' structure of Episode 3}

We now turn to show the ``nested''  structure of the late X-ray luminosity. \citet{Pisani2013} have shown that the X-ray rest-frame $0.3$--$10$ keV luminosity light curves present a constant decreasing power-law behavior, at $t_a\gtrsim10^4$ s, with typical slopes of $-1.7\lesssim\alpha_X\lesssim-1.3$. 
This has been proven in a sample of six BdHNe: GRBs 060729, 061007, 080319B, 090618, 091127, and 111228, hereafter \textit{golden sample} \citep[GS, see, e.g.,][]{IzzoRel,Pisani2013}.
That the late X-ray emission could play a fundamental role as a distance indicator has been explored inferring the redshifts of GRBs 101023 and 110709B \citep{Penacchioni2011,Penacchioni2013}.
The IGC paradigm also allowed predicting $\sim10$--$15$ days in the cosmological rest frame before its discovery, the occurrence of the SN associated to GRB 130427A, the most luminous source ever observed in $\gamma$ rays with $E_{iso}\approx10^{54}$ erg and $z=0.34$ \citep{2013GCN..14478...1X,2013GCN..14491...1F}.
This was later confirmed by the observations \citep{Postigo2013cq,2013GCN..14686...1L,2013GCN..14666...1W,GCN2013conf}.

We compare and contrast GRB 130427A X-ray data with GRB 060729, a member of the GS, and GRB 061121, which shows the general behavior of BdHNe.
GRB 060729, at $z=0.54$, has $E_{iso}=1.6\times10^{52}$ erg \citep{Grupe2007b} and a SN bump in its optical afterglow \citep{Cano2010}. 
GRB 061121, at $z=1.314$ \citep{2006GCN..5826....1B}, has $E_{iso}=3.0\times10^{53}$ erg, and its Episode 4 is clearly missing in view of the high cosmological redshift.

\begin{figure}
\centering
\includegraphics[width=0.7\hsize,clip]{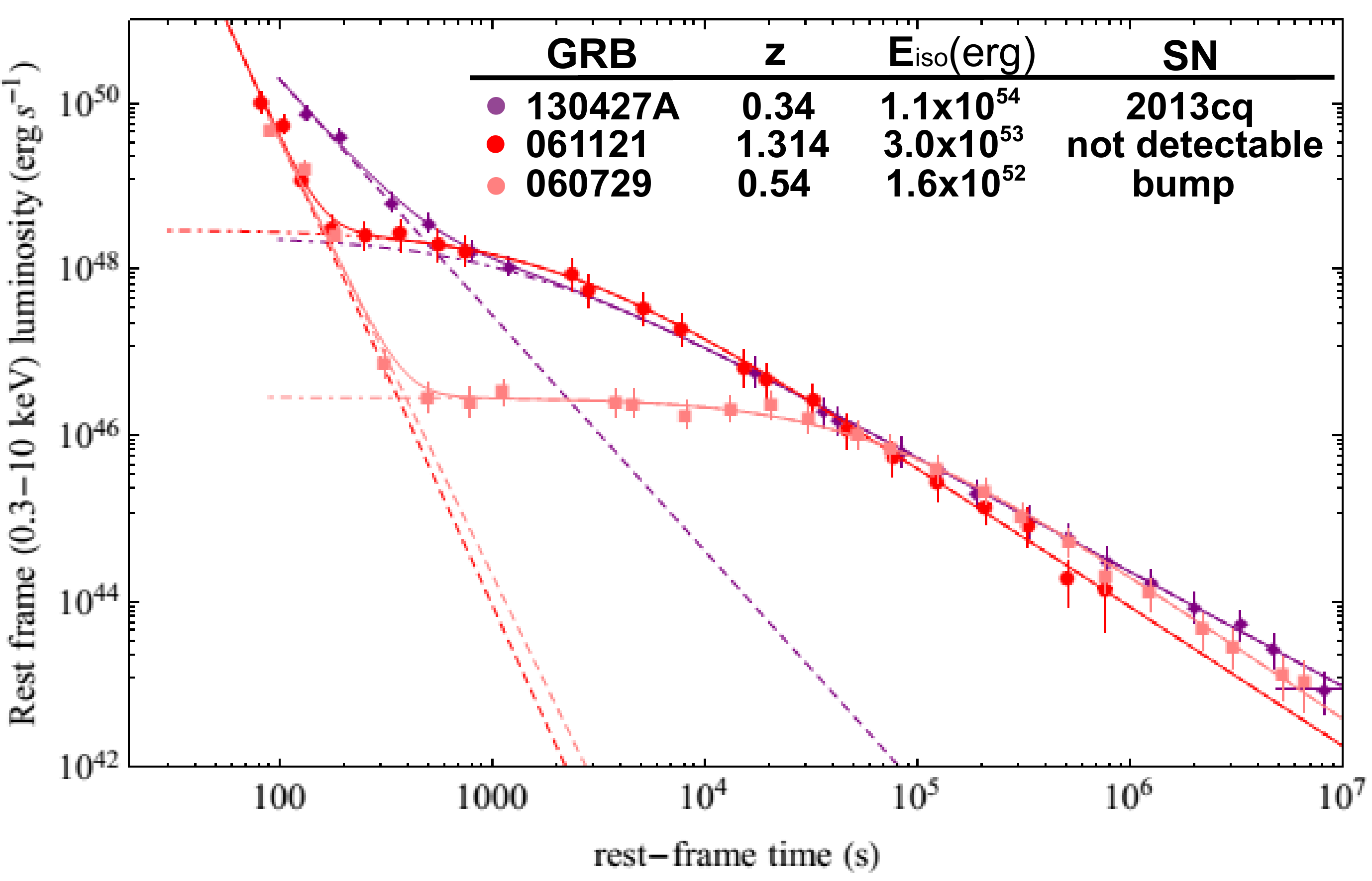}
\caption{Rest-frame $0.3$--$10$ keV re-binned luminosity light curves of GRB 130427A (purple), GRB 061121 (red, shifted by $50$s in rest frame), and GRB 060729 (pink). The light curves are fitted by using a power-law for the steep decay phase (dashed lines) and the function in Eq. (\ref{plateaulatedecay}) for the plateau and the late decay phases (dot-dashed curves).}
\label{family_GRB130427A}
\end{figure}

In Fig. \ref{family_GRB130427A} we have plotted the rebinned rest-frame $0.3$--$10$ keV luminosity light curves of GRBs 130427A, 060729, and 061121. Their steep decay is modeled by a power-law function, i.e. $L_p\left(t_a/100\right)^{-\alpha_p}$, where $L_p$ and $\alpha_p$ are the power-law parameters.
The plateau and the late power-law decay are instead modeled by using the following phenomenological function
\begin{equation}
\label{plateaulatedecay}L(t_a)=L_X (1+t_a/\tau)^{\alpha_X}\, ,
\end{equation}
where $L_X$, $\alpha_X$, and $\tau$, respectively, are the plateau luminosity, the late power-law decay index, and the characteristic timescale of the end of the plateau.
From Eq. (\ref{plateaulatedecay}), we have defined the end of the plateau at the rest-frame time $t^\ast_a=\tau[(1/2)^{1/\alpha_X}-1]$, when the luminosity of the plateau is half of the initial one, $L_a(t^\ast_a)=L_X/2$.

From this fitting procedure, we can conclude that the three BdHN systems considered here share the following properties:\\
\textbf{a)} the power-law decay for the more energetic sources starts directly from the steep decay, well before the $t_a\approx2\times10^4$ s, as indicated in \citet{Pisani2013}. Consequently, the plateau shrinks as a function of the increasing $E_{iso}$ (see Fig. \ref{family_GRB130427A});\\
\textbf{b)} the luminosities in the power-law decay are uniquely functions of the cosmological rest-frame arrival time $t_a$ independently on the $E_{iso}$ of each source (see Fig. \ref{family_GRB130427A});\\
\textbf{c)} most remarkably, the overlapping of the X-ray light curves reveals a ``nested'' structure of BdHN Episodes 3.

\begin{figure*}
\centering
\hfill
\includegraphics[width=0.358\hsize,clip]{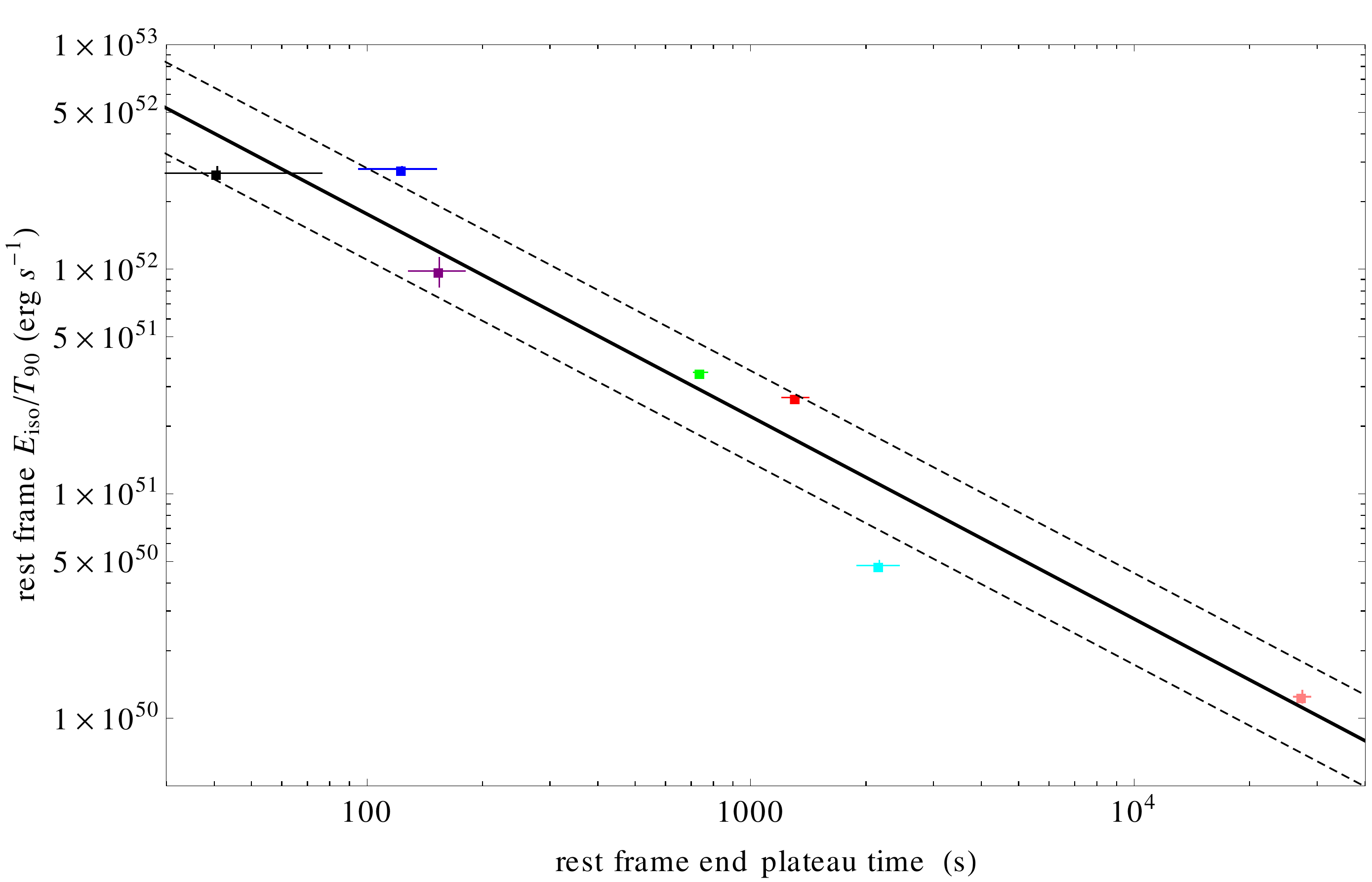}
\hfill
\includegraphics[width=0.345\hsize,clip]{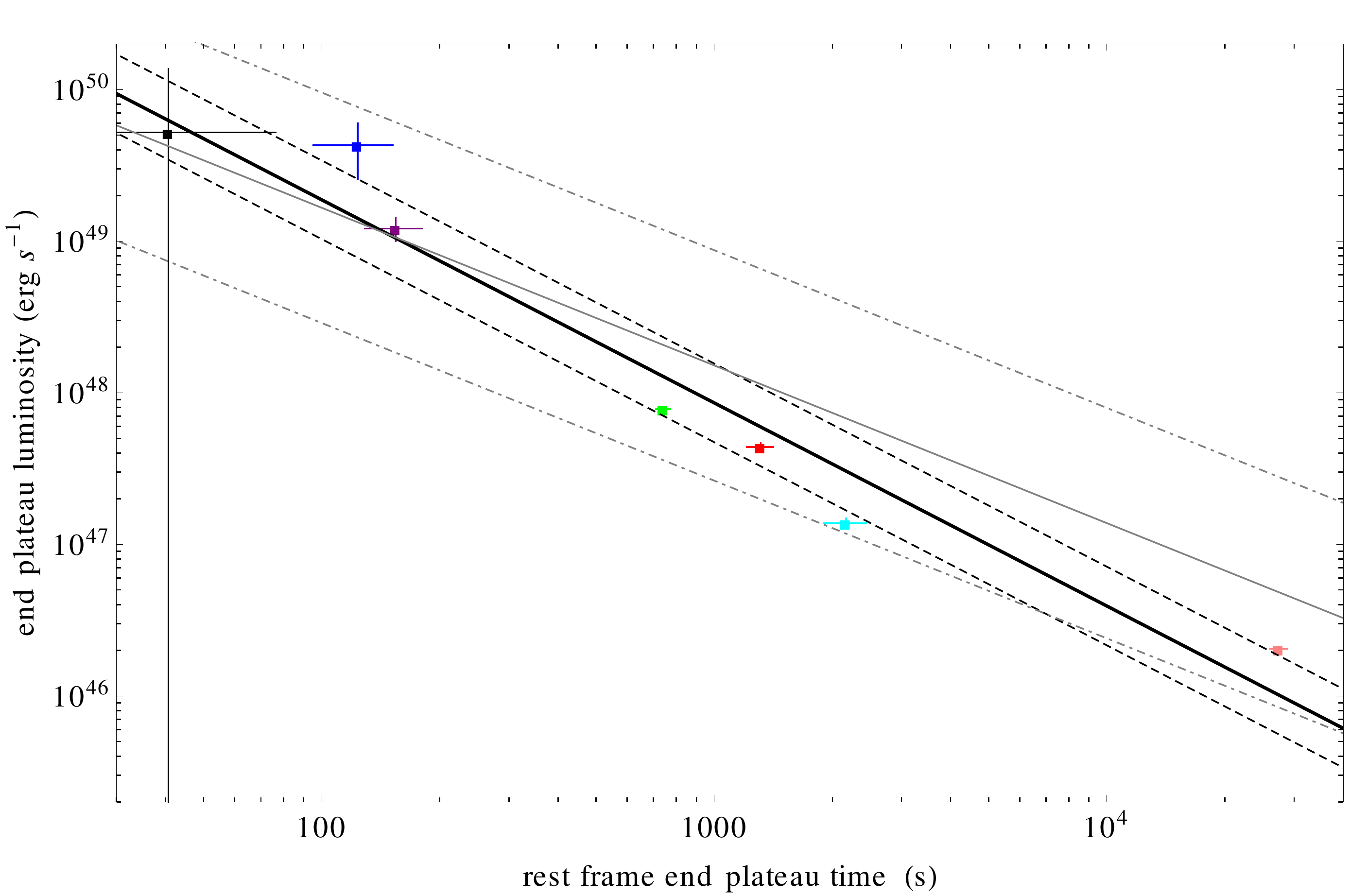}
\hfill\null
\caption{The $\langle L_{iso}\rangle$--$t^\ast_a$ (left) and the $L_a$--$t^\ast_a$ (right) correlations (solid black lines) and the corresponding $1\sigma$ confidence levels (dashed black lines). The considered sources are GRB 060729 (pink), GRB 061007 (black), GRB 080319B (blue), GRB 090618 (green), GRB 091127 (red), GRB 111228A (cyan), and GRB 130427A (purple). The tigher BdHNe $L_a$--$t^\ast_a$ correlation is compared to the one in \citet{Dainotti2011b}, corresponding to $m=-1.04$ and $q=51.30$ (solid gray line) and $\sigma=0.76$ (dotted-dashed gray lines).}
\label{correlations}
\end{figure*}

In our sample of BdHNe, we verify the applicability of the Dainotti-Willingale relations $\langle L_{iso}\rangle$--$t^\ast_a$ and $L_a$--$t^\ast_a$ \citep{Dainotti2008,Dainotti2011,Willingale2007}, where $\langle L_{iso}\rangle=E_{iso}/t_{a,90}$ is the averaged luminosity of the prompt and $t_{a,90}$ is the rest-frame $t_{90}$ duration of the burst.
The resulting correlations, $\log_{10}Y_i=m_i \log_{10}X_i+q_i$, are shown in Fig. \ref{correlations}.
The parameters of each BdHN and the best fit parameters, $m_i$ and $q_i$ (where $i=1$,$2$), are summarized in Table \ref{corrs}.
As is clear from the extra scatter values $\sigma_i$, our total BdHN sample provides tighter correlations. The extra scatter of the $L_a$--$t^\ast_a$, $\sigma=0.26$, is less than the \citet{Dainotti2011b} ones, i.e., $\sigma=0.76$ for the whole sample of $62$ bursts and $\sigma=0.40$ for the best subsample of eight bursts ($U0095$).
The Dainotti-Willingale correlations consider X-ray afterglows characterized by a steep decay, a plateau phase, and a late power-law decay \citep{Nousek2006,Zhang2006}, independently of their energetics. 
In our BdHN sample we limit the attention to a) the most energetic sources, $10^{52}$--$10^{54}$ erg, b) the presence of four emission episodes (neglecting Episode 4 for $z>1$), and c) sources with determined redshift and complete data at $t_a=10^4$--$10^6$ s.
All these conditions appear to be necessary to fulfill the nested structure in Fig. \ref{family_GRB130427A} and the tighter correlations between the astrophysical parameters $\langle L_{iso}\rangle$, $L_a$, and $t^\ast_a$ in Fig. \ref{correlations}.

\begin{table}
\centering
\caption{List of the quantities of the considered sources and best fit parameters of the correlations in Fig. \ref{correlations}.}
{\tiny
\begin{tabular}{cccc}
\hline\hline
\textbf{GRB}  &  \textbf{$\langle L_{iso}\rangle$ ($10^{50}$erg/s)}  &  \textbf{$t^\ast_a$ (ks)}  &  \textbf{$L_a$ ($10^{47}$erg/s)}                                    \\
\hline
060729        &  $1.25\pm0.08$                                       &  $27.4\pm1.4$              &  $0.20\pm0.01$\\
061007        &  $267\pm18$                                          &  $0.041\pm0.036$           & $521\pm\textnormal{unc}$   \\
080319B       &  $279\pm7$                                           &  $0.12\pm0.03$             &  $430\pm170$  \\
090618        &  $34.7\pm0.3$                                        &  $0.74\pm0.03$             &  $7.81\pm0.17$\\
091127        &  $26.8\pm0.3$                                        &  $1.31\pm0.10$             &  $4.39\pm0.26$\\
111228A       &  $4.79\pm0.24$                                       &  $2.17\pm0.27$             &  $1.38\pm0.10$\\
130427A       &  $98\pm15$                                           &  $0.16\pm0.03$                &  $121\pm21$\\
\hline\hline
\textbf{Correlation}                   &  \textbf{$m_i$}    &  \textbf{$q_i$} &  \textbf{$\sigma_i$}  \\
\hline
$\langle L_{iso}\rangle$--$t^\ast_a$    &  $-(0.90\pm0.09)$  &  $54.0\pm0.3$   &  $0.20\pm0.05$        \\
$L_a$--$t^\ast_a$                       &  $-(1.34\pm0.14)$  &  $52.0\pm0.4$   &  $0.26\pm0.08$        \\
\hline
\end{tabular}
}
\label{corrs}
\end{table}

To explain the above nested power-law decay and constrained correlations, we consider the decay of heavy elements produced in the r-process as a viable energy source \citep{BBFH1957}, originating in binary NS mergers \citep[see, e.g.,][]{Li1998,Janka1999,Rosswog2004,Oechslin2007,Goriely2011,Piran_r_process}.

\citet{Li1998} have shown that the emission from the surface of an optically thick expanding ejecta in an adiabatic regime provides a flat light curve \citep[see also][]{Arnett1982}. 
This can explain, in principle, the observed steep decay and plateau phase of Episode 3 (see Fig. \ref{family_GRB130427A}). 
After the ejecta becomes transparent, the heating source term due to the nuclear decays of the heavy nuclei, generated via r-process, becomes directly observable and dominates. 
The avalanche of decays with different lifetimes then provides the total energy release per unit mass per time that follows a power-law distribution, whose decay index has been estimated to be $-1.4\lesssim\alpha\lesssim-1.1$ \citep{Metzger2010}.
These values are strikingly similar to the ones we have found in the late X-ray luminosity.

This power-law behavior is different from the exponential decay observed in the optical light curves of traditional SN, powered by the decay of a single element ($^{56}$Ni $\to$ $^{56}$Co $\to$ $^{56}$Fe), which is not produced in the avalanche of many decays as in the r-process.

\section{Conclusions}

To summarize, short GRBs have been shown to come from binary NS mergers (see, e.g., \citealp{Goodman1986,Paczynski1986,Eichler1989,MeszarosRees1997_b,Rosswog2003,Lee2004}; and more recently \citealp{Muccino2013}). Our subclass of long, extremely energetic ($10^{52}$--$10^{54}$ erg) sources is also initially driven by a tight binary system, formed by a $\nu$-NS and a companion NS, surrounded by the SN ejecta (see Fig. \ref{schema}). Then we denoted these most energetic GRBs by ``BdHNe''. This is clearly different from the gravitational collapse of a single massive progenitor star described by the collapsar model \citep{Woosley1993,MacFadyenWoosley,2006ARA&A..44..507W}.

We compared and contrasted the late X-ray luminosities of three BdHNe with different $E_{iso}$, finding a nested structure. We showed tight correlations between $\langle L_{iso}\rangle$, $L_a$ and $t^\ast_a$ (see Fig. \ref{correlations} and Table \ref{corrs}) in agreement with the Dainotti-Willingale ones.

The above scaling laws, the nesting, and the initial dimension of $\sim7\times10^{12}$ cm and Lorentz factor of $\Gamma\approx2$ obtained from the steep decay of the X-ray luminosity put stringent limits on alternative theoretical models. They do not appear to be explainable within the traditional fireball jetted model, originating in the synchrotron radiation emitted by a decelerating relativistic shell with $\Gamma\sim10^2$ and colliding with the circumburst medium at distances $\sim10^{16}$ cm \citep[see, e.g.,][and reference therein]{Sari1998b,Piran2005,Meszaros2006,Gehrels2009}. In this Letter we alternatively proposed that the late X-ray luminosity comes from the wide angle emission of the SN ejecta or in the accretion on the newly born BH. We call the attention on the role of the energy release in the SN ejecta from the decay of very heavy nuclei generated by r-process in binary NSs \citep{Li1998}. This heavy nuclei avalanche decay \citep[see, e.g.,][]{Metzger2010} may well explain the late X-ray luminosity of Episode 3. This emission follows the steep decay and plateau phase of the adiabatic optically thick expansion, prior to reaching transparency (see Fig. \ref{family_GRB130427A}).

In the case of binary systems with longer periods and/or a lower accretion rate, which do not allow the NS companion to reach its critical mass and to form a BH, Episode 2 is missing. 
The presence of the companion NS will neverthless strip the H and He envelopes of the core progenitor star.
These sources have low energetic bursts ($E_{iso}<10^{52}$ erg), such as GRB 060218 and GRB 980425, and their X-ray luminosity light curves do not overlap with the ones of our more energetic sample of BdHNe.
These systems do not conform to the IGC paradigm and are traditional hypernovae\footnote{http://nsm.utdallas.edu/texas2013/proceedings/3/1/Ruffini.pdf}.

\begin{acknowledgement}
This work made use of data supplied by the UK Swift Science Data Center at the University of Leicester.
ME, MK, and GBP are supported by the Erasmus Mundus Joint Doctorate Program by grant Nos. 2012-1710, 2013-1471, and 2011-1640, respectively, from the EACEA of the European Commission.
We warmly thank the anonymous referee for very constructive suggestions that improved the presentation of this Letter.
\end{acknowledgement}

\end{document}